\documentclass[%
 reprint,
superscriptaddress,
showpacs,preprintnumbers,
 amsmath,amssymb,
 aps,
]{revtex4-1}
\usepackage{braket}
\usepackage{soul}
\usepackage{ulem}
\usepackage{graphicx}
\usepackage{dcolumn}
\usepackage{bm}
\usepackage{color}
\newcommand\redout{\bgroup\markoverwith{\textcolor{red}{\rule[.5ex]{2pt}{0.4pt}}}\ULon}

\usepackage[colorlinks=true,citecolor=blue]{hyperref}
\hypersetup{colorlinks=true,citecolor=blue,linkcolor=red,urlcolor=blue}

\begin{document}

\title{ Resonance spin transfer torque in ferromagnetic/normal/ferromagnetic spin-valve structure of topological insulators }

\author{Moslem Zare }
\affiliation{Department of Physics, Yasouj University, Yasouj, Iran 75914-353, Iran.}

\begin{abstract}
We theoretically study the spin current and spin-transfer torque generation in a conventional spin-valve hybrid structure of type ferromagnetic/normal metal/ferromagnetic (FM/NM/FM) made of the topological insulator (TI), in which a gate voltage is attached to the normal layer. We demonstrate the penetration of the spin-transfer torque into the right ferromagnetic layer and show that, unlike graphene spin-valve junction, the spin-transfer torque in TI is very sensitive to the chemical potential of the NM region.
As an important result, by changing the chemical potential of the NM spacer and magnetization directions, one can control all components of the STT. Interestingly, both the resonance spin current and the resonance spin-transfer torque appear for energies determined from a resonance equation.
By increasing the chemical potential of the NM spacer, the amplitude of the STTs decreases while at large chemical potentials of $\mu_N$ there are intervals of chemical potential in which both the spin current and the spin-transfer torque become zero.
These findings could open new perspectives for applications in spin-transfer torque magnetic random access memory (STT-MRAM) devices based on TI.
 \end{abstract}

\maketitle

\section{Introduction}
The electric current modulation of the magnetic properties of magnetic materials instead of externally applied magnetic fields has paved the way to integrate magnetic functionalities into electric-current-controlled spintronics devices with reduced dimensions and energy consumption compared with conventional magnetic field actuation.
The conservation of angular momentum between itinerant electrons and localized magnetization in magnetic heterostructures leads to the of particular interest concept of spin-transfer torque (STT)~\cite{Berger,Slonczewski}, plays a major role in spintronic devices ~\cite{Tsoi98,Katine2000,Huai2004,Diao2005}. In this phenomenon, the spin angular momentum of electrons in a spin-polarized current, generated by passing an electrical current through a ferromagnet layer, exerts a torque on the second magnetization, enabling magnetization switching or precession \cite{Ralph2008,Brataas12}, for sufficiently large currents without the need for an external field. It is found to be important because of its potential for applications in spin-torque diode effect~\cite{Tulapurkar}, microwave-assisted recording of hard-disk drives~\cite{Zhu2008,Zhu2010}, high-performance, and high-density magnetic storage devices~\cite{fabian,Bauer12nat,Lindernat15}.
Compared with current memory devices which use magnetic fields to reorient magnetization to store information ,Spin-transfer torque magnetic random access memory (STT-MRAM) devices, which store information in the magnetization of a nanoscale magnet, is a promising candidate for the last two decades
~\cite{Ralph2008,Brataas12,Kent15,Apalkov16,Chung16}. Magnetic-nonmagnetic multilayers such as magnetic tunnel junctions, spin valves, point contacts, nanopillars, and nanowires~\cite{Brataas12} are common structures and device geometries that are applicable for STT proposal.
Among them, as originally proposed ~\cite{Tang95,Miyazaki}, magnetic tunnel junctions were used as a high-performance, non-volatile magnetic memory cells in MRAMs~\cite{Parkin99}. As a large current is needed for current-induced magnetization dynamics, for creating the current densities required for the onset of magnetic instabilities ($10^{8} A/cm^{2}$) nanometre-scale devices should be used.
Despite the explosive growth of the field of STT in three-dimensional materials, only a few works have studied the spin-transfer torque of two-dimensional heterostructures. STT generation in ferromagnetic-normal-ferromagnetic bulk graphene junctions has been studied theoretically in Ref. ~\cite{Yokoyama11}, then possibility of current-induced STT in ferromagnetic-normal-ferromagnetic graphene nanoribbon junction studied by Ding et al.~\cite{Ding14}. Very recently in a detailed study, we theoretically investigated the transport and STT in phosphorene-based multilayers with noncollinear magnetizations\cite{moslem-stt,moslem-physicac}.
In the present work, motivated by the recent measurements of the STT induced by a topological insulator~\cite{Mellnik14}, we theoretically study the generation of the spin currents and STT in F/N/F trilayer heterostructures of TI. Within the scattering formalism, we find that the application of a local gate voltage to the N region of the FM/NM/FM structure leads to both the spin current and the spin-transfer torque resonance. Depending on the chemical potential of the NM region ($\mu_N$), and the configuration of the magnetization vectors one can has STTs.
\begin{figure}[h]
\centering
\includegraphics[width=8.4cm]{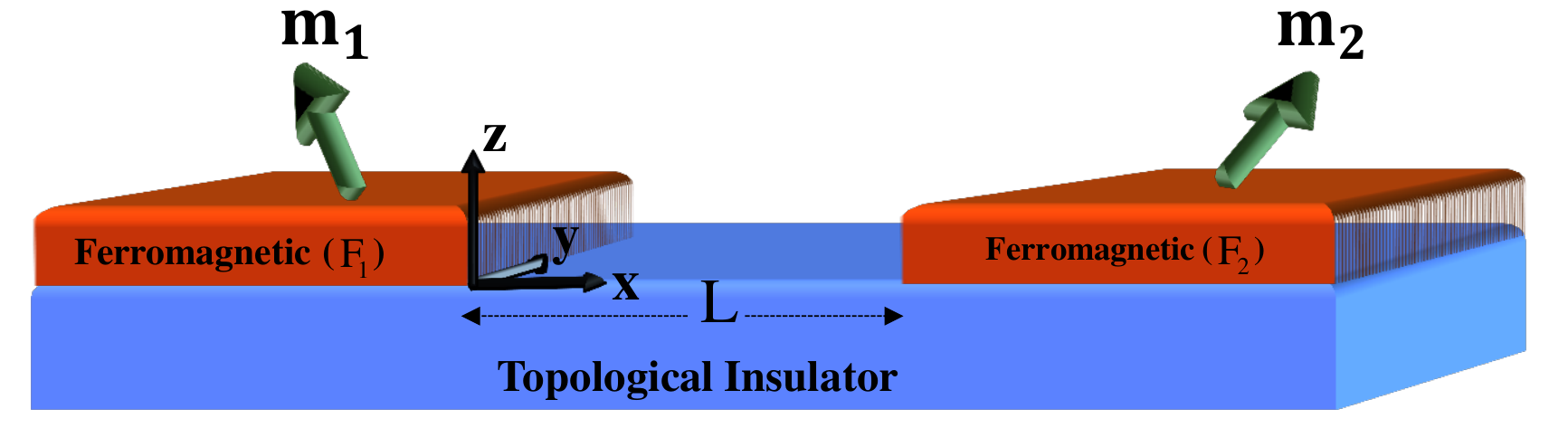}
\caption{\label{fig:1} (Color online) Schematic illustration of a FM-TI/NM-TI/FM-TI heterostructure, where the total charge current is flowing along the x axis through the left ferromagnetic layer ($F_1$) to the right ferromagnetic one ($F_2$). The green arrows represent the local magnetic moments with overall magnetization directions ${\bf m}_1,{\bf m}_2$.}
\end{figure}

This paper is organized in the following way: In Sec. \ref{MODEL}, we introduce the low-energy effective Hamiltonian of the ferromagnetic topological insulator and establish the theoretical framework which is used to calculate the spin current and spin-transfer torque generation in a conventional spin-valve hybrid structure of type ferromagnetic/normal metal/ferromagnetic (FM/NM/FM) made of the topological insulator (TI), in which a gate voltage is attached to the normal layer. In Sec. \ref{sec:Nsumerical}, we discuss our numerical results for the proposed FM/NM/FM hetrostructure. Finally, our conclusions are summarized in Sec. \ref{summary}.

\section{MODEL AND BASIC FORMALISM}  \label{MODEL}
As illustrated in Fig. \ref{fig:1}, we consider a conventional spin-valve hybrid structure of type ferromagnetic ($F_1$)/normal metal (NM)/ferromagnetic ($F_2$) made of the topological insulator, with a normal spacer of width $L$. In general ${\bf m}_1$ and ${\bf m}_2$ in which ${\bf m}=(m_x,m_y,m_z)=|{\bf m}|(\sin\theta\cos\phi,\sin\theta\sin\phi,\cos\theta)$ are vectors along the magnetization of the left and right layers, respectively which are uniform and can point along any general direction. $\theta, \phi$ denote polar and azimuthal angles in the spherical coordinate, respectively. Flowing a current from layer $F_1$ into region $F_2$  induces a spin accumulation in $NM/F_2$ interface, exserted a spin-transfer torque on the magnetization $F_2$~\cite{Berger,Slonczewski}. We suppose a normal metal spacer much thinner than the spin relaxation length in TI. It is worth mentioning that in order to realize ferromagnetic TI, one may utilize either doping the TI with magnetic impurities \cite{Fan14,Henk12,Chen10} or using the proximity effect by coating it with a ferromagnetic insulator\cite{scholz2012,Mellnik14}.

The low-energy effective Hamiltonian of the ferromagnetic TI near the Dirac point, can be written as \cite{Vaezisot}
\begin{eqnarray} \label{H0}
\hat H=\hbar v\hat{\boldsymbol{\sigma}}\cdot({\bf k}\times{\bf z})+ \hat{\boldsymbol{\sigma}}\cdot{\bf m}-\varepsilon_{\rm F},
\end{eqnarray}
where $\hat{\boldsymbol{\sigma}}$ and $v$ are the spin space Pauli matrices and the Fermi velocity, respectively and ${\bf z}$ denotes the unit vector in the $z$ direction. For simplicity, hereafter we set $\hbar v=1$
The electron transport is confined in the $x$-$y$ plane and ${\bf k} =(k_x,k_y,0)=k(\cos\phi_k,\sin\phi_k,0)$. The second term in Eq. \eqref{H0} is the exchange coupling between itinerant and local spins and $\varepsilon_{\rm F}$ is the Fermi energy.\par
\begin{center}
\begin{figure}[h]
\centering
\includegraphics[width=1.56in]{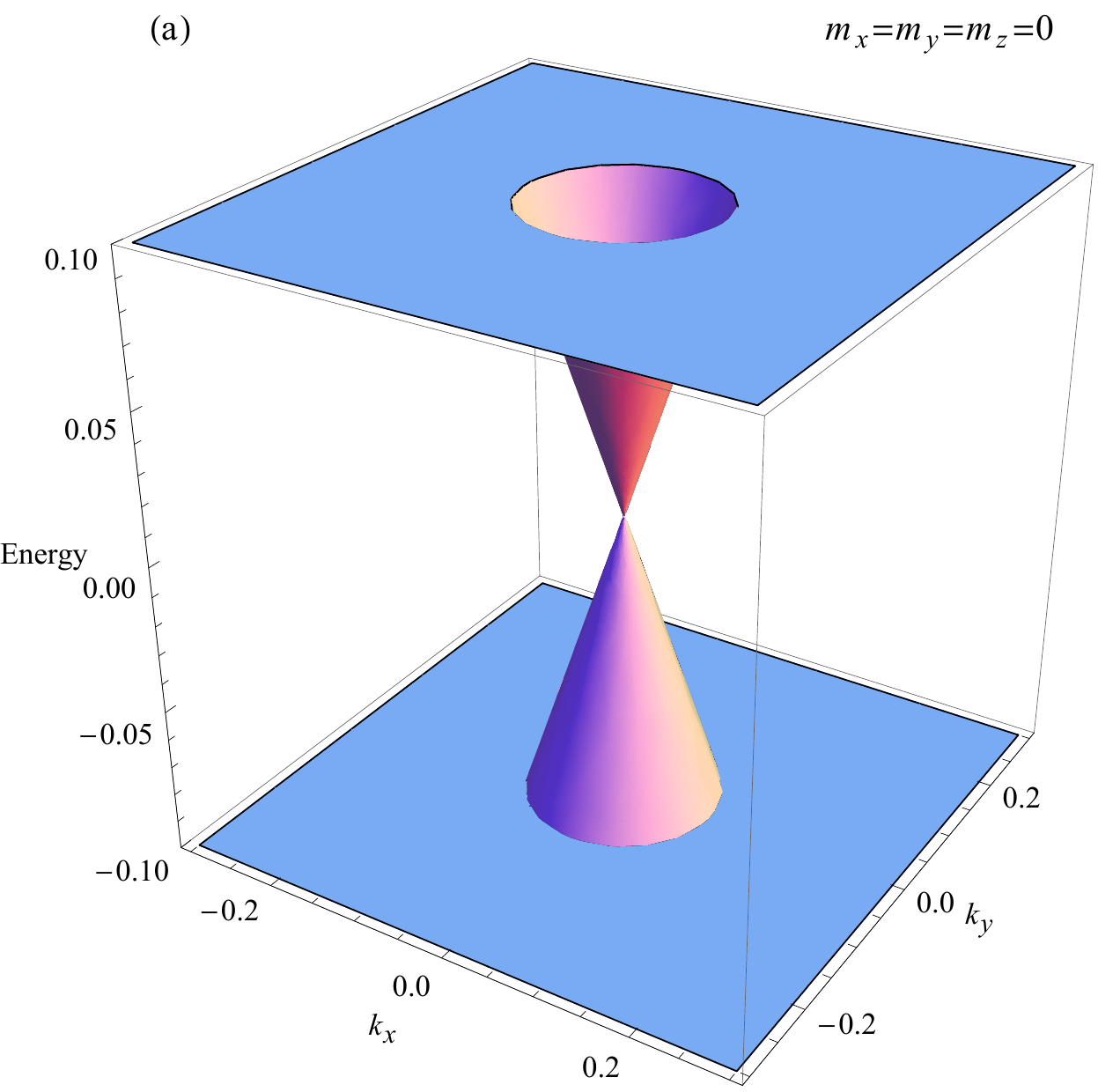}.
\includegraphics[width=1.56in]{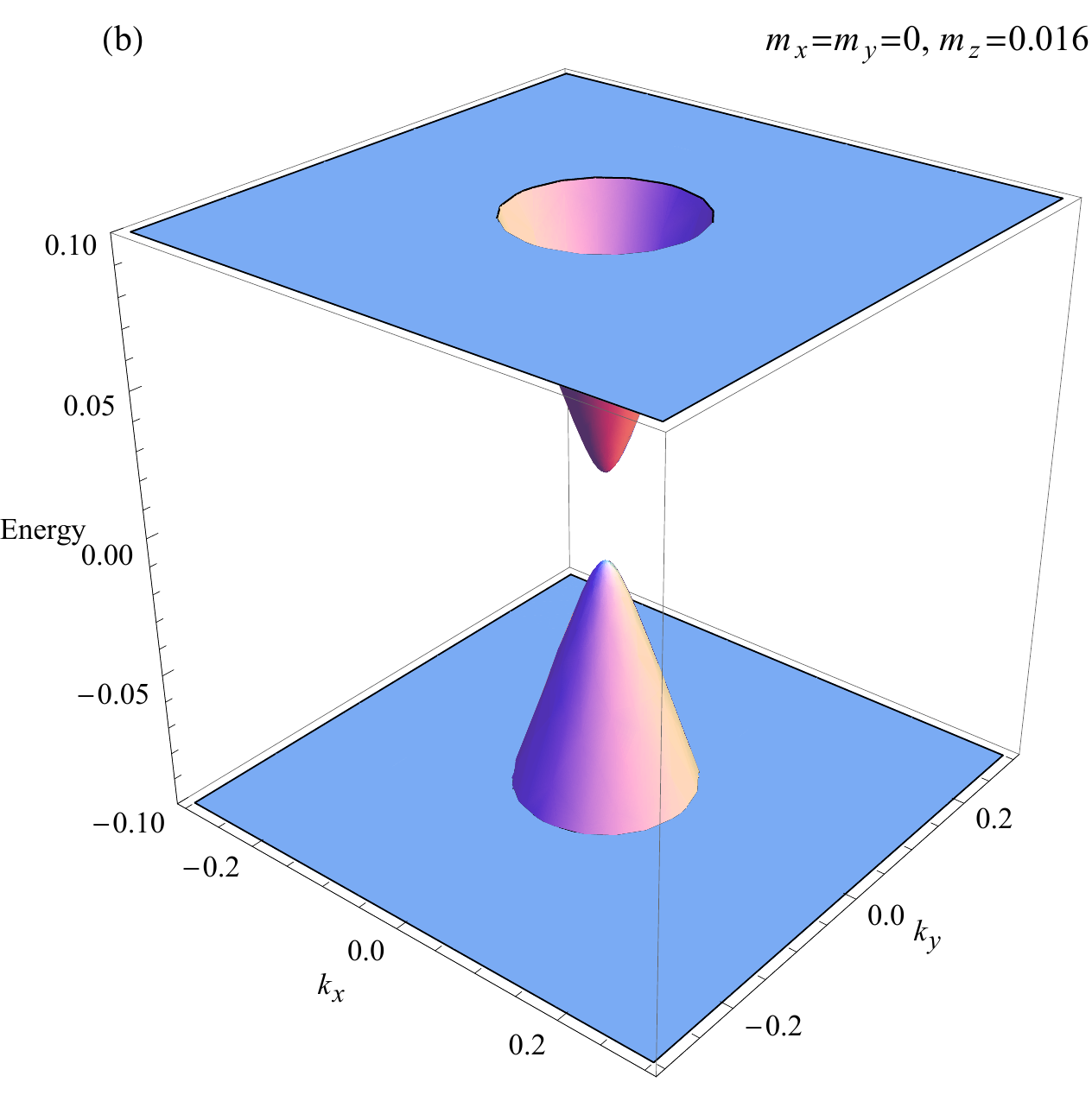}.
\includegraphics[width=1.56in]{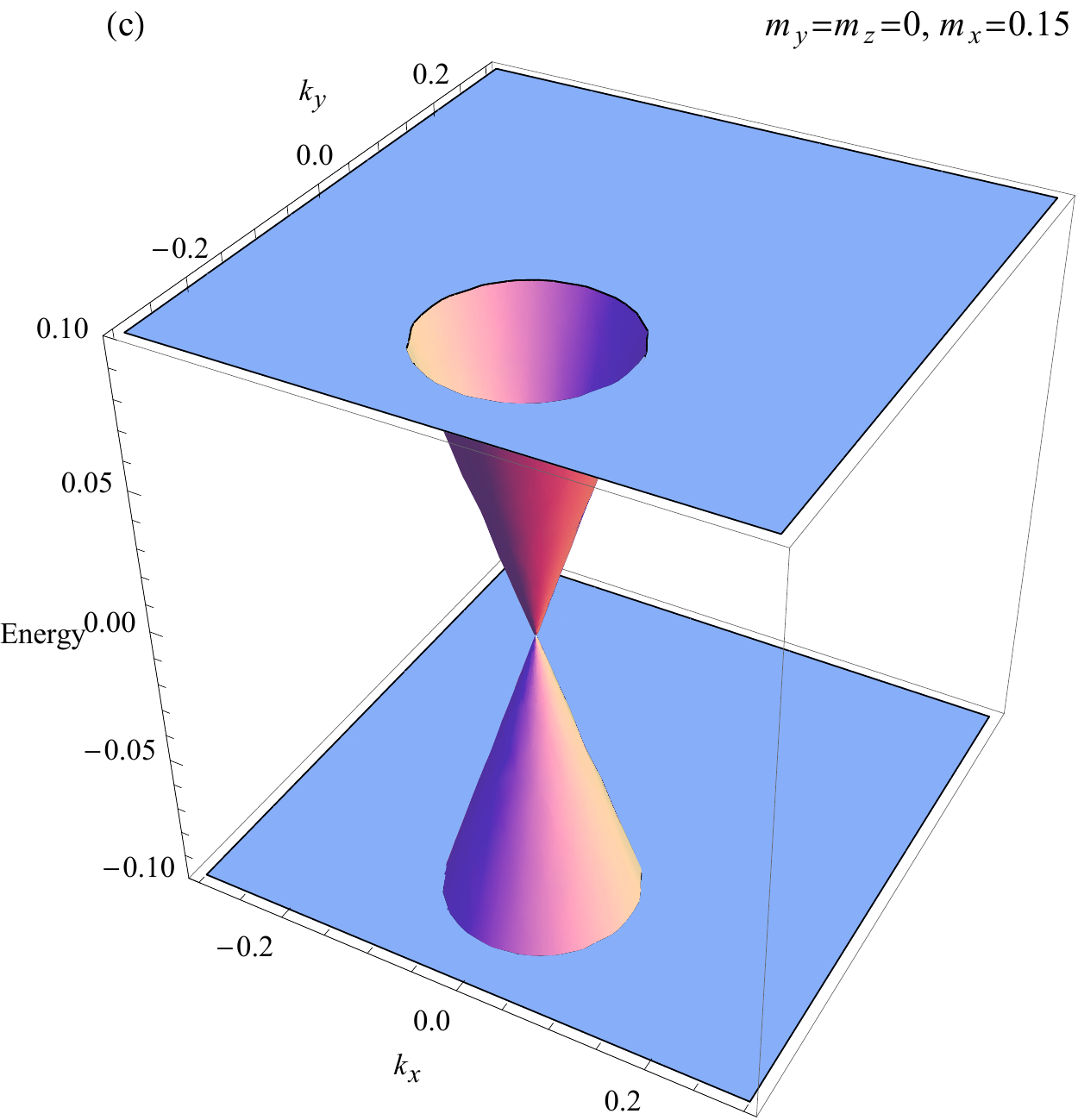}.
\includegraphics[width=1.56in]{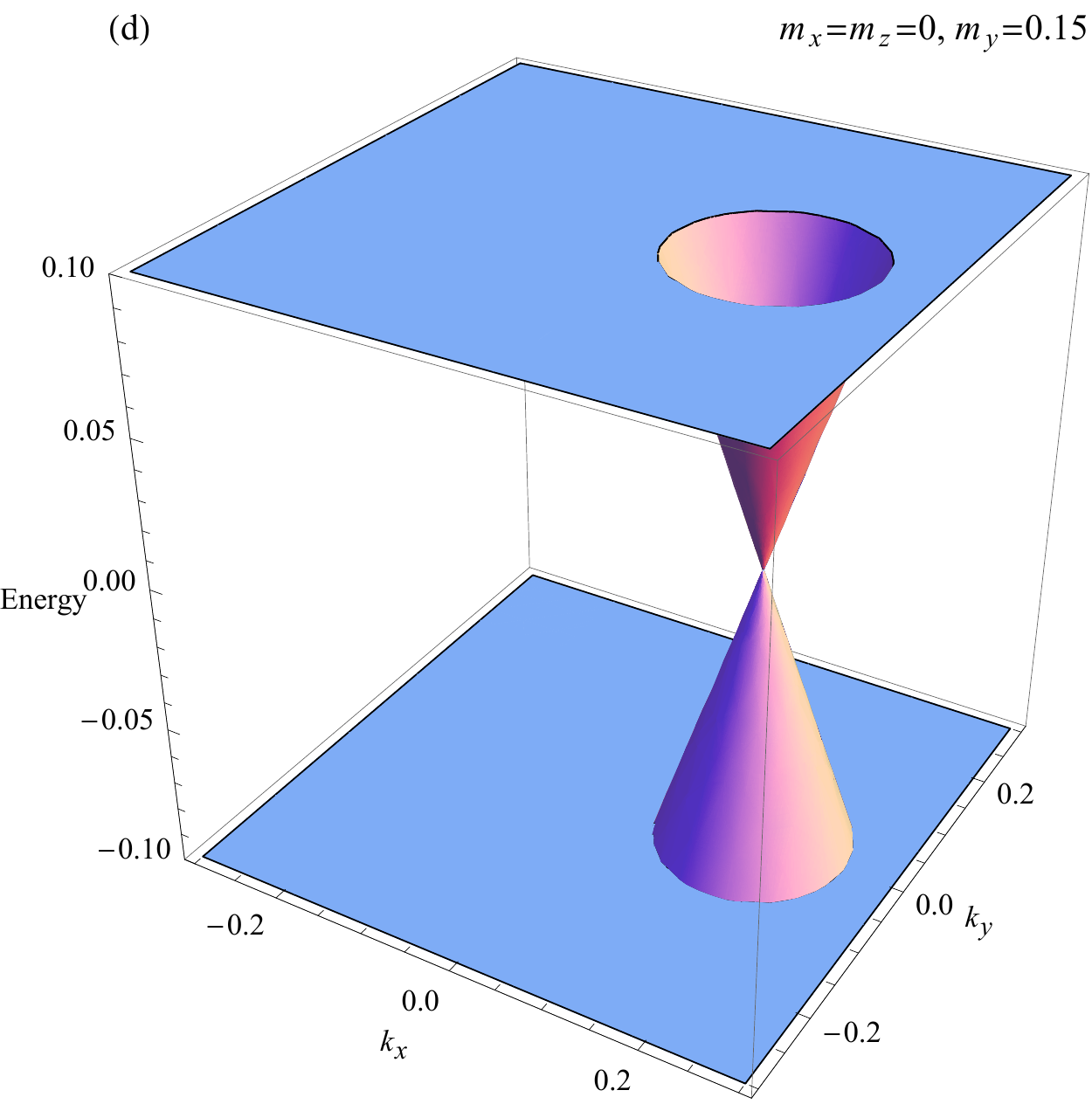}.
\caption{(Color online) The band structure of the topological insulator (a)in the absence of the magnetization  ($m_x=m_y=m_z=0$), (b) $m_x=m_y=0,m_z=0.016$, (c) $m_y=m_z=0,m_x=0.15$ and (d) $m_x=m_z=0,m_y=0.15$, in the presence of the magnetization.
\label{Fig:dis}}
\end{figure}
\end{center}

Figure \ref{Fig:dis} shows the band structure of the pristine and ferromagnetic TI, (a) $m_x=m_y=m_z=0$, (b) $m_x=m_y=0,m_z=0.016$, (c) $m_y=m_z=0,m_x=0.15$ and (d) $m_x=m_z=0,m_y=0.15$. In the absence of the magnetization (the case of (a)), the band structure consists of a massless Dirac cone. An energy gap can be opened in the spectrum of TI when the magnetization lies out of plane ($m_z\neq 0$ (b)). It is worth mentioning that the $x$ and $y$ components of the magnetization, have no effect on the gap modification of the TI band structure and only shift the Dirac cone along the $y$ and $x$-momentum axis, respectively (c,d).
\onecolumngrid
\twocolumngrid
The wave functions that diagonalize the unperturbed Dirac Hamiltonian $\hat{H}$ (Eqn.\ref{H0}) are explicitly given as
\begin{equation}\label{eq:chiralbasis}
\ket{u_+^{F}}=\begin{pmatrix}e^{i\alpha_k}\cos\frac{\beta_k}{2}\\ \sin\frac{\beta_k}{2}\end{pmatrix},\;\ket{u_{-}^{F}}=\begin{pmatrix}-e^{i\alpha_k}\sin\frac{\beta_k}{2}\\\cos\frac{\beta_k}{2}\end{pmatrix},
\end{equation}
with
$\alpha_k=\tan^{-1}\left [ \frac{\hbar vk\cos\phi_k-m\sin\phi\sin\theta}{\hbar vk\sin\phi_k+m\cos\phi\sin\theta} \right ],\;\beta_k=\cos^{-1}\left [\frac{m}{|\varepsilon_{\bf k}|}\cos\theta \right ],$ and $\varepsilon_{k}^{\pm}=\pm\sqrt{\hbar^2v^2k^2+m^2+2\hbar vkm\sin\theta\sin (\phi_k-\phi)}$.
For the normal region the spinors are as
\begin{eqnarray}\label{eq:chiralbasis}
\ket{u_{\pm}^{N}}=\begin{pmatrix}\pm e^{i\alpha^N_k}\\1 \end{pmatrix},
\end{eqnarray}

in which, $\alpha^N_k=\tan^{-1}\left [ \frac{k_x}{k_y} \right ]$.
When a spin-polarized current interacts with a ferromagnetic layer due to the spin filtering, a spin transfer torque is applied to the magnetic layer. Supposing that there is no spin-flipping processes, overall transmission and reflection amplitudes for spin-up electrons ($t_{\uparrow}$, $r_{\uparrow}$) are different from those of spin-down electrons ($t_{\downarrow}$, $r_{\downarrow}$).
Total wave functions in the two ferromagnetic regions are as
\begin{eqnarray}
\label{eq:wavefunction1}
\psi^{F_1}_{\rm in} = {e^{ik_x x} \over \sqrt{{\Omega}}}
\Big( e^{i\alpha^+_k}\cos(\beta_k/2) \left|
\uparrow \right\rangle
+\sin(\beta_k/2) \left|\downarrow \right\rangle \Big).
\end{eqnarray}
\begin{eqnarray}
\label{eq:wavefunction2}
\psi^{F_1}_{\rm ref} = {e^{-ik_x x} \over \sqrt{{\Omega}}}
\Big( r_{\uparrow}e^{i\alpha^-_k}\cos(\beta_k/2) \left|
\uparrow \right\rangle
+r_{\downarrow}\sin(\beta_k/2) \left|\downarrow \right\rangle \Big).
\end{eqnarray}
\begin{eqnarray}
\label{eq:wavefunction3}
\psi^{F_2}_{\rm tran} = {e^{ik_x x} \over \sqrt{{\Omega}}}
\Big( t_{\uparrow}e^{i\alpha^+_k}\cos(\beta_k/2) \left|
\uparrow \right\rangle
+t_{\downarrow}\sin(\beta_k/2) \left|\downarrow \right\rangle \Big).
\end{eqnarray}

The corresponding eigenvectors in the normal region can be written as
\begin{eqnarray}
\label{eq:wavefunction4}
\psi^{\pm N} = {e^{\pm ik^N_x x} \over \sqrt{{2\Omega}}}
\Big( a e^{i\alpha_k^{\pm N}} \left|
\uparrow \right\rangle
+b \left|\downarrow \right\rangle \Big)
\end{eqnarray}

Here, $\alpha_k^{\pm N}=\alpha_k^N(\pm \phi_k)$ and ${\Omega}$ is a normalization area. The two propagation directions along the $x$ axis are denoted by $\pm$ in $\Psi^{\pm N}$. By matching the wave functions and their first derivatives at the interfaces $x = 0$
and $x = L$, we obtain the coefficients in the wave functions as
\begin{widetext}
\begin{eqnarray}
\label{tud}
r_{\uparrow}
&=& \frac{e^{i(\alpha^+_{1k}-\alpha^-_{1k})}(e^{2ik_x^{N}L}(k_ {x}^{F2}-k_x^{N})(k_ {x}^{F1}+k_x^{N})-(k_ {x}^{F1}-k_x^{N})(k_ {x}^{F2}+k_x^{N})}{e^{2ik_x^{N}L}(k_x^{N}-k_ {x}^{F1})(k_x^{N}-k_ {x}^{F2})-(k_x^{N}+k_ {x}^{F1})(k_x^{N}+k_ {x}^{F2})}\nonumber\\
r_{\downarrow}
&=&- \frac{e^{2ik_x^{N}L}(k_x^{N}+k_ {x}^{F1})(k_x^{N}-k_ {x}^{F2})+(k_ {x}^{F1}-k_x^{N})(k_ {x}^{F2}+k_x^{N})}{e^{2ik_x^{N}L}(k_x^{N}-k_ {x}^{F1})(k_x^{N}-k_ {x}^{F2})-(k_x^{N}+k_ {x}^{F1})(k_x^{N}+k_{x}^{F2})}
\end{eqnarray}
\begin{eqnarray}
\label{tud}
t_{\uparrow}
&=& \frac{e^{i(\alpha^+_{1k}-\alpha^+_{2k}+(k_x^{N}-k_ {x}^{F_2})L
)}k_ {x}^{F1}k_x^{N}\cos(\beta_{1k}/2)\sec(\beta_{2k}/2)}{e^{2ik_x^{N}L}(k_x^{N}-k_ {x}^{F1})(k_x^{N}-k_ {x}^{F2})-(k_x^{N}+k_ {x}^{F1})(k_x^{N}+k_ {x}^{F2})}\nonumber\\
t_{\downarrow}
&=& -\frac{e^{i(k_x^{N}-k_ {x}^{F_2})L}k_ {x}^{F1}k_x^{N}\sin(\beta_{1k}/2)\csc(\beta_{2k}/2)}{e^{2ik_x^{N}L}(k_x^{N}-k_ {x}^{F1})(k_x^{N}-k_ {x}^{F2})-(k_x^{N}+k_ {x}^{F1})(k_x^{N}+k_ {x}^{F2})}
\end{eqnarray}
\end{widetext}
In the steady state, the spin transfer torque acting on a volume $V$ of material (by conservation of angular momentum) can be
computed simply by determining the net flux of non-equilibrium spin current ${\bf J^S}$ through the surfaces of that volume as
\begin{eqnarray}
\label{js0}
\label{eq:tau}
               {\bf \tau}_{\rm stt} =- \int_{V} dV \nabla
\cdot {\bf J^S},
\end{eqnarray}
Note that since ${\bf J^S}$ is a tensor, its dot product with a vector in real space leaves a vector in spin space.
For a single-electron wavefunction $\psi$, similar to the more-familiar probability current density $(\hbar/m){\rm
Im}(\psi^*{\bf \nabla} \psi)$, the spin current density can be rewritten as
\begin{eqnarray}
\label{eq:js2}
             {\bf J^S_{ij}} = {\hbar \over m} {\rm Im}
             (\psi^* {\bf S_i} \otimes {\partial_j}\psi),
\end{eqnarray}
Here $i, j=x,y$, with $i$ indicating the spin component and $j$ the
transport direction. $m$ is the electron mass, and
${\bf S}$ represents the Pauli matrices $S_x$, $S_y$,
and $S_z$. The three spin current density components can be determined substituting
Eqs.~(\ref{eq:wavefunction1}-\ref{eq:wavefunction4}) into Eq.\ref{eq:js2} as
\begin{eqnarray}
\label{eq:qs}
J^S_{xx(y),trans} & = & {\hbar^2 k \over 2m{\Omega}} 2 {\rm Re({\rm Im})}[t_{\uparrow}e^{i\alpha^+_{2k}}\cos(\beta_{2k}/2)t_{\downarrow}\sin(\beta_{2k}/2)] \nonumber \\
J^S_{xz,trans} & = & {\hbar^2 k \over 2m{\Omega}} [|t_{\uparrow}|^2\cos^2(\beta_{2k}/2) - |t_{\downarrow}|^2 \sin^2(\beta_{2k}/2)]. \nonumber
\end{eqnarray}
\begin{eqnarray}
\label{eq:qs}
J^S_{xx(y),in} & = & {\hbar^2 k \over 2m{\Omega}} 2 {\rm Re({\rm Im})}[e^{i\alpha^+_{1k}}\cos(\beta_{1k}/2)\sin(\beta_{1k}/2)] \nonumber \\
J^S_{xz,in} & = & {\hbar^2 k \over 2m{\Omega}} \cos(\beta_{1k}). \nonumber
\end{eqnarray}
\begin{eqnarray}
\label{eq:qs}
J^S_{xx(y),ref} & = & {\hbar^2 k \over 2m{\Omega}} 2 {\rm Re({\rm Im})}[r_{\uparrow}e^{i\alpha^-_{1k}}\cos(\beta_{1k}/2)r_{\downarrow}\sin(\beta_{1k}/2)] \nonumber \\
J^S_{xz,ref} & = & {\hbar^2 k \over 2m{\Omega}} [|r_{\uparrow}|^2\cos^2(\beta_{1k}/2) - |r_{\downarrow}|^2 \sin^2(\beta_{1k}/2)]. \nonumber
\end{eqnarray}
It is clear that the total spin current is not conserved during
the filtering process because the spin current density flowing on the
left of the magnet ${\bf J^S_{\rm in}} + {\bf J^S_{\rm refl}}$ is not
equal to the spin current density on the right ${\bf J^S_{\rm trans}} $.
Using Eq.~(\ref{js0}), the spin transfer torque
${\bf \tau}_{\rm stt}$ on an area $A$ of the ferromagnet is equal to the net spin current
transferred from the electron to the ferromagnet, and is given by ${\bf \tau}_{\rm stt}= A {\bf \hat{x}} \cdot ({\bf J^S_{\rm in}} + {\bf J^S_{\rm refl}}- {\bf J^S_{\rm trans}})$.
Using the scattering theory as well as the incoherency of spin-up and -down states inside the ferromagnet, the STT can be formulated in terms of the spin dependence of the transmission and reflection coefficients as

\begin{eqnarray}
                   { \tau^{x(y)}_{\rm st}}
                   &=& {A \over \Omega}{\hbar^2 k \over m} {\rm Re({\rm Im})}\Big[k_1\cos(\beta_{1k}/2)\sin(\beta_{1k}/2)(e^{i\alpha^+_{1k}}\nonumber\\
                  &-&r_{\uparrow}r_{\downarrow}^*e^{i\alpha^-_{1k}}) -k_2 t_{\uparrow}t_{\downarrow}^*e^{i\alpha^+_{2k}}\cos(\beta_{2k}/2)\sin(\beta_{2k}/2)  \Big]
\nonumber\\
                 { \tau^z_{\rm st}}
                    &=& {A \over \Omega}{\hbar^2 k \over m} \Big[|t_{\uparrow}|^2(k_1\cos^2(\beta_{1k}/2)-k_2\cos^2(\beta_{2k}/2))\nonumber\\
                    &-&|t_{\downarrow}|^2(k_1\sin^2(\beta_{1k}/2)-k_2\sin^2(\beta_{2k}/2))  \Big]
\label{eq:Model1}
\end{eqnarray}
We have used the fact that $|t_{\uparrow}|^2 + |r_{\uparrow}|^2 = 1$ and
$|t_{\downarrow}|^2 + |r_{\downarrow}|^2 = 1$. It is worth mentioning that for a symmetric F/N/F junction there is no component of spin torque in the ${\bf \hat{z}}$ direction and the other two components are as follow
\begin{eqnarray}
                   { \tau^{x(y)}_{\rm st}}
                   &=& {A \over \Omega}{\hbar^2 k \over 2m} \sin(\beta_k) {\rm Re({\rm Im})}\Big[e^{i\alpha^+_k}(1-t_{\uparrow}t_{\downarrow}^*) - e^{i\alpha^-_k} r_{\uparrow}r_{\downarrow}^* \Big]\nonumber\\
\label{eq:Model1}
\end{eqnarray}
By including all transverse modes, the total STT of the proposed structure at zero temperature is given by
\begin{equation}
\label{conductance}
\tau_{tot}^i (E) =\int_{0}^{k_{y}^{max}(E)} \tau^i(E,k_y)\ dk_y,
\end{equation}
$k_{y}^{max}(E)$ is the maximum value of the transverse momentum.
\section{Nsumerical results}\label{sec:Nsumerical}
In this section, we present our numerical results. As described in the introduction, a conventional spin-valve structure has the general ferromagnetic/normal/ferromagnetic structure. We study the generation of the spin currents and the spin-transfer torque in a spin-valve hybrid structure of type topological insulator junctions.
As we are interested in both the metallic (where the chemical potential stands away from the charge neutrality point) and the zero energy regime, a gate voltage is attached to the normal layer. We set $m_1=m_2=m=0.1$ eV in all figures and results presented in this section.
\onecolumngrid
\begin{center}
\begin{figure}[h]
\includegraphics[scale=0.45]{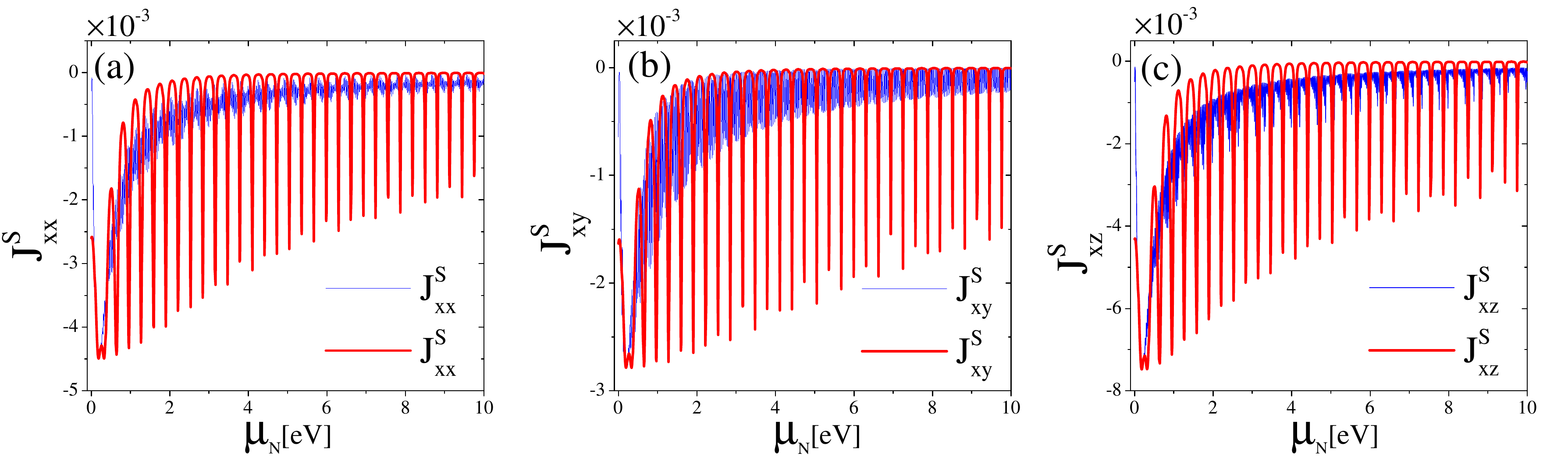}
\caption{(Color online) The transmitted spin current density (in units of ${\hbar^2}/{m\Omega}$) versus the chemical potential of the NM spacer ($\mu_N$), for when the first and second magnetizations are fixed along the $z$ $(\theta_1=0)$ and $-z$ $(\theta_2=\pi)$ axes, respectively. (a)${J}_{xx}^S$ (b) ${J}_{xy}^S$  (c) ${J}_{xz}^S$. The red curves are for $L=10$ and the blue ones are for $L=100$. The other parameters are taken as $m=0.1$ eV, $\mu_F=0.2$ eV.
\label{Fig:0}}
\end{figure}
\end{center}
\twocolumngrid
Figure \ref{Fig:0} shows the transmitted spin current densities (a)${J}_{xx}^S$ (b) ${J}_{xy}^S$  (c) ${J}_{xz}^S$ (in units of ${\hbar^2}/{m\Omega}$) versus the chemical potential of the NM spacer ($\mu_N$), when the first and second magnetizations are fixed along the $z$ $(\theta_1=0)$ and $-z$ $(\theta_2=\pi)$ axes, respectively. The red curves are for $L=10$ and the blue ones are for $L=100$. The other parameters are taken as $m=0.1$ eV, $\mu_F=0.2$ eV. As can be seen, the spin current density is an oscillatory function of the chemical potential of the NM spacer and amplitude of the oscillations drops with increasing the chemical potential of the NM region. The spin current density of a junction with a thicker normal region exhibits faster oscillations.
Interestingly, resonant spin current peaks appear at the chemical potentials of the NM region that satisfy the equation $k_x^NL=2n\pi$ with $n$ the positive integer where $k_x^N$ and $L$ are the wavevector and width of the NM region, respectively.
\onecolumngrid
\begin{center}
\begin{figure}[h]
\includegraphics[scale=0.75]{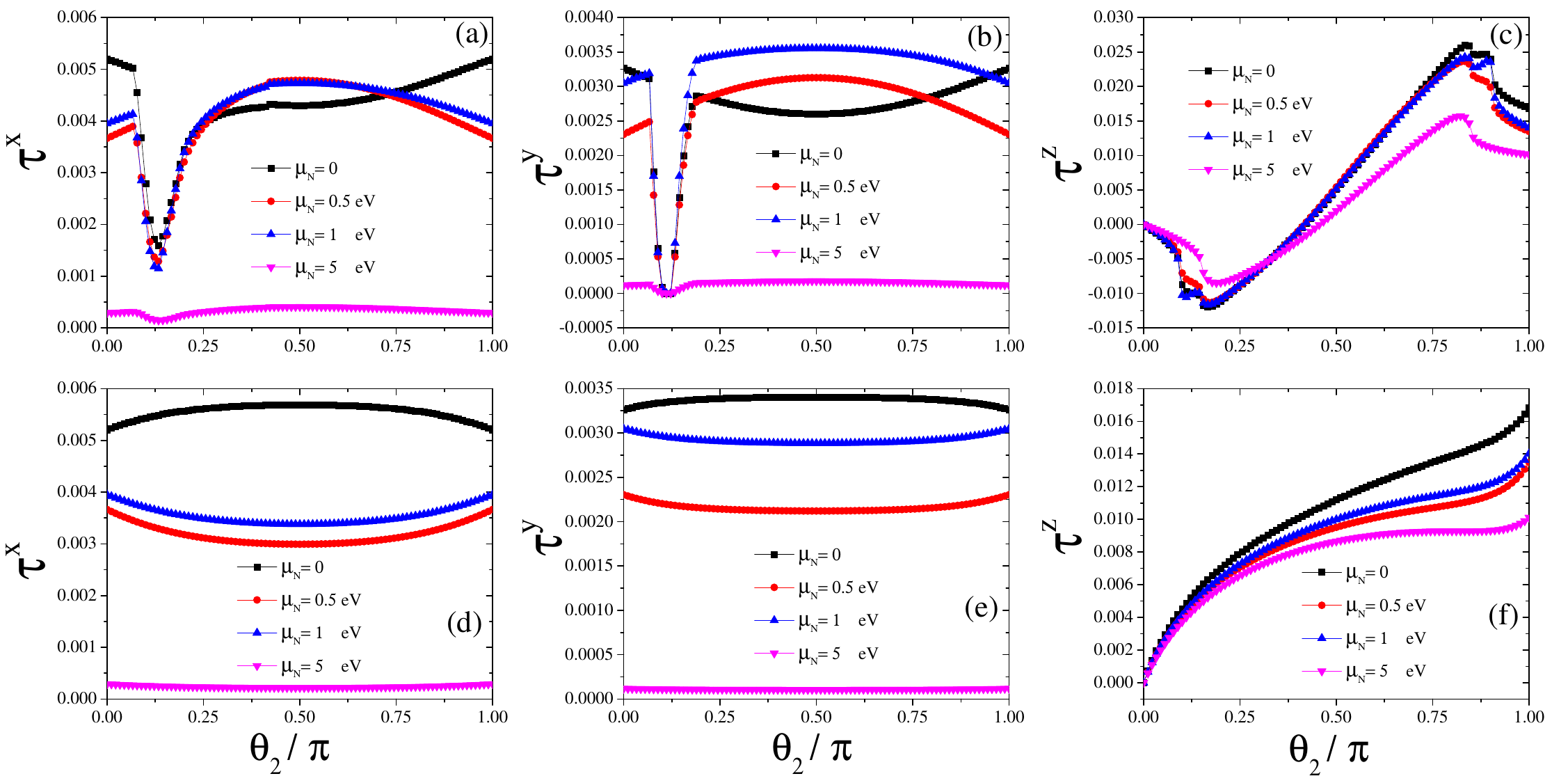}
\caption{(Color online) The spin-transfer torques (in units of ${A\hbar^2}/{m\Omega}$) versus the polar angle of the second magnetization vector $\bf{m}_2$ ($\theta_2$) for various chemical potential of the NM region ($\mu_N$). (a,d) ${\tau}_{STT}^x$ and (b,e) ${\tau}_{STT}^y$  and (c,f) ${\tau}_{STT}^z$. Top and bottom panels are for $\phi_{1(2)}=0(\pi)$ and $\phi_{1(2)}=\pi/2(3\pi/2)$, respectively. The other parameters are taken as $\theta_1=0$, $\mu_F=0.2$ eV, $m=0.1$ eV and $L=10$.
\label{Fig:1}}
\end{figure}
\end{center}
\twocolumngrid
The dependence of the STT components on the polar angle of the second magnetization ($\theta_2$), for various chemical potentials of the NM region $\mu_N$ is presented in Fig.\ref{Fig:1}, (a,d) ${\tau}_{STT}^x$ and (b,e) ${\tau}_{STT}^y$ and (c,f) ${\tau}_{STT}^z$. Top and bottom panels are for $\phi_{1(2)}=0(\pi)$ and $\phi_{1(2)}=\pi/2(3\pi/2)$, respectively. The other parameters in this figure are taken as $\theta_1=0$, $\mu_F=0.2$ eV, $m=0.1$ eV and $L=10$. In both panels the first magnetization fixed along the $z$-axis. The second magnetization rotates from the $z$-axis to the $-x$-axis, (inside the $x$-$z$ plane) in the top panel and from the $z$-axis to the $-y$-axis, (inside the $y-z$ plane) in the bottom panel. As a whole, we see that the STT components decrease with increasing the chemical potential of the NM region. Maximum STTs are related to the zero gate voltage. For the configuration $(\theta_{1(2)},\phi_{1(2)})=(0(\theta_2),0(\pi))$, in a certain angle, STTs reached to the maximum value.
At the configuration $(\theta_{1(2)},\phi_{1(2)})=(0(\theta_2),\pi/2(3\pi/2))$, ${\tau}_{STT}^x$ and ${\tau}_{STT}^y$ are symmetric for the interval $[0,\pi]$. The $z$ component of the STT (${\tau}_{STT}^z$) in the parallel configuration ($\theta_1=\theta_2=0$), for each chemical potential $\mu_N$ becomes zero. The $x, y$ components of the STT reach their maximum value at $\theta_2=\pi/2$ (parallel configuration), while the ${\tau}_{STT}^z$ component obtains its maximum at $\theta_2=\pi$ (antiparallel configuration).
\begin{center}
\begin{figure}[h]
\includegraphics[scale=0.35]{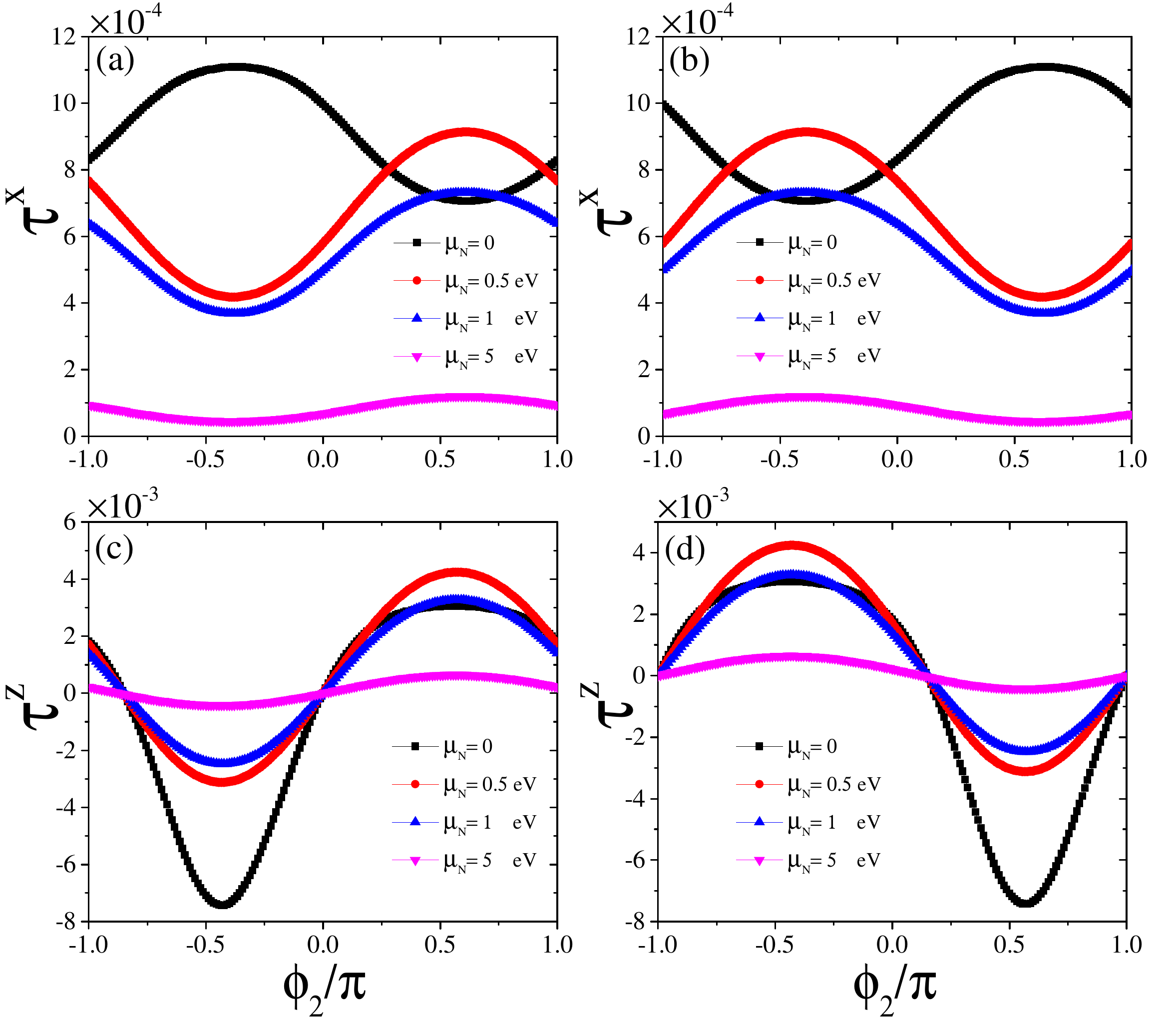}
\caption{(Color online) The spin-transfer torques (in units of ${A\hbar^2}/{m\Omega}$) versus the azimuthal angle of the second magnetization vector $\bf{m}_2$ ($\phi_2$), for various chemical potential of the NM spacer ($\mu_N$).(a,b) ${\tau}_{STT}^x$ (c,d) ${\tau}_{STT}^z$. Left (a,c) and right (b,d) panels are for $\theta_{1(2)}=\pi/2(\pi/2)$ and $\theta_{1(2)}=\pi/2(3\pi/2)$, respectively. The other parameters are taken as $\phi_1=0$, $\mu_F=0.2$ eV, $m=0.1$ eV and $L=10$. In both configurations, the $y$-component of the STT (${\tau}_{STT}^y$) becomes zero.
\label{Fig:2}}
\end{figure}
\end{center}
The dependence of the STTs on the azimuthal angle, for different values of the chemical potential of the NM region $(\mu_N)$, is shown in Fig.\ref{Fig:2}. As the sign tunability in the STTs devices is crucial, we see that one can simply control STTs both in terms of sign and magnitude. As seen, regardless of the magnetization configuration, the dependence on the azimuthal angle essentially follows the usual sinusoidal behavior, in agreement with Ref.~\cite{Mellnik14}. Except $\mu_N=0$, the oscillations amplitude of the STTs decreases with increasing chemical potential of the normal TI. STTs of these two configurations have a phase difference of 180 degrees. In both configurations, the case of $\mu_N=0$ has a phase shift of 180 degrees in ${\tau}_{STT}^x$, relative to other chemical potentials of the NM region.
\onecolumngrid
\begin{center}
\begin{figure}[h]
\includegraphics[scale=0.45]{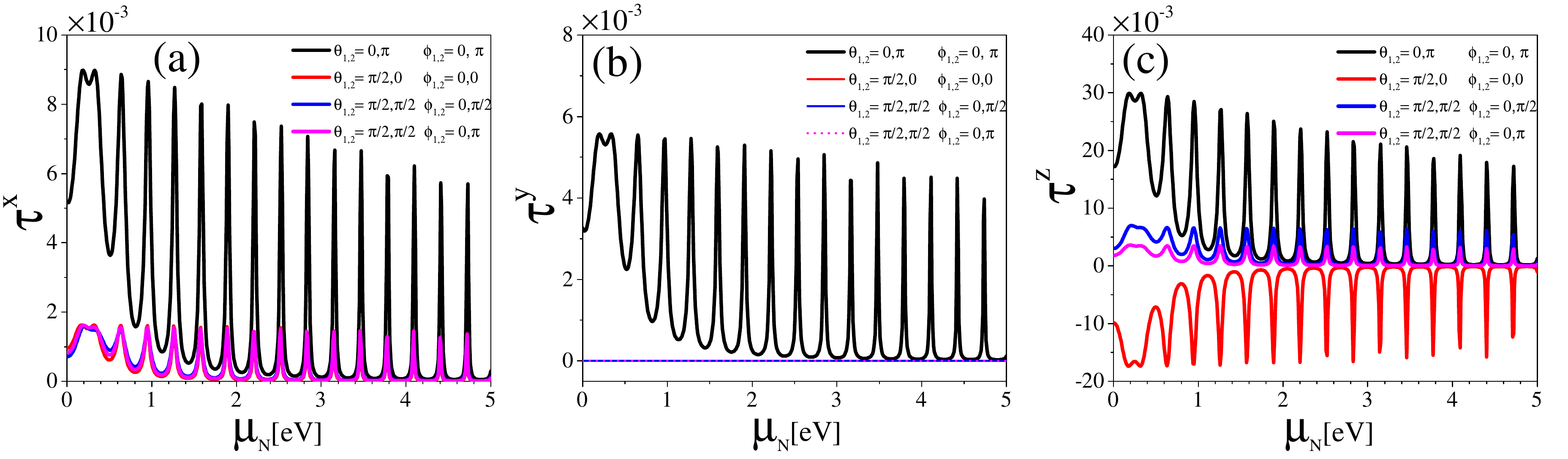}
\caption{(Color online) The spin-transfer torques (a) ${\tau}_{STT}^x$ and (b) ${\tau}_{STT}^y$ and (c) ${\tau}_{STT}^z$ (in units of ${A\hbar^2}/{m\Omega}$) versus the chemical potential of the NM spacer ($\mu_N$), for various configurations of the magnetizations. The other parameters are taken as $\mu_F=0.2$ eV, $m=0.1$ eV and $L=10$.
\label{Fig:3}}
\end{figure}
\end{center}
\begin{center}
\begin{figure}[h]
\includegraphics[scale=0.45]{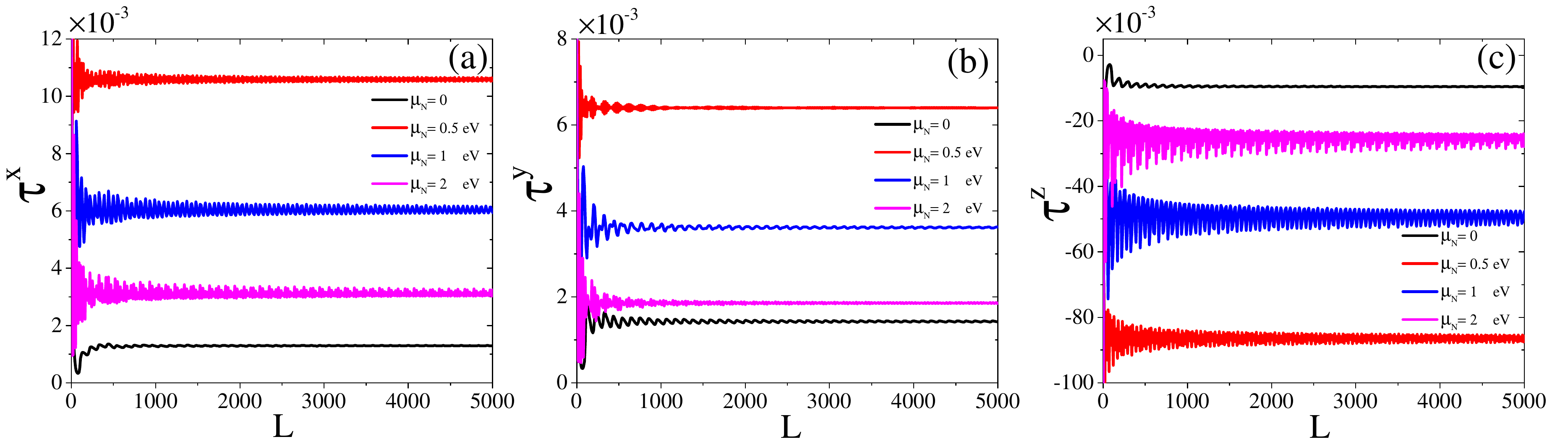}
\caption{(Color online) The spin-transfer torques (a) ${\tau}_{STT}^x$ and (b) ${\tau}_{STT}^y$ and (c) ${\tau}_{STT}^z$ (in units of ${A\hbar^2}/{m\Omega}$) versus the length of the NM region ($L$), for various chemical potential of the NM spacer ($\mu_N$). The left ($\bf{m}_1$) and right ($\bf{m}_2$) magnetizations are fixed along the $z$ and $y$-axis, respectively.
\label{Fig:4}}
\end{figure}
\end{center}
\twocolumngrid
In Figure \ref{Fig:3}, we show the effect of chemical potential of the NM region ($\mu_N$) on the spin-transfer torque components, (a,d) ${\tau}_{STT}^x$ and (b,e) ${\tau}_{STT}^y$  and (c,f) ${\tau}_{STT}^z$. The results are shown for different configurations of the magnetizations $\bf{m}_1$ and $\bf{m}_2$. The other parameters are taken as $\mu_F=0.2$ eV, $m=0.1$ eV and $L=10$.
As an important result, we see that by changing $\mu_N$, one can control the STT that could be a useful consequence for the applications in TI-based nano-electronic devices. Also, note that the amplitude of the STT oscillations decreases as the chemical potential of the NM region increases. Furthermore, the formation of resonant-STT in the right ferromagnetic region is achievable by changing the $\mu_N$.
In excellent agreement with Ref.~\cite{Zhang17}, a clear oscillatory behavior with sharp peaks in STTs is observed . It is also found that more peaks appear in a same Fermi energy region with enhancing the width of the NM spacer. Increasing the chemical potential of the NM region leads to the STT resonance occurs at values of the energies determined from the equation $k_x^NL=2n\pi$. Interestingly, all of the components of the STT are symmetric with respect to the sign reversal of the chemical potential.
It is further seen that at large chemical potentials $\mu_N$ there are intervals of potential in which the spin-transfer torques become zero.

In Figure \ref{Fig:4}, we plot the spin-transfer torque versus the thickness of the central NM layer, when the first and second magnetizations fixed along the $y$ and $z$-axis, respectively. The STTs display rapid oscillations as a function of normal TI (spacer) width. It is easily seen that for each configuration, the amplitude of the STT oscillations decays with the spacer width. The magnitude of the STTs oscillations decreases as the chemical potential of the NM region increases.

\section{summary}\label{summary}
In summary, we theoretically study the spin current and spin-transfer torque generation in a conventional spin-valve hybrid structure of type ferromagnetic/normal metal/ferromagnetic (FM/NM/FM) made of the topological insulator (TI), in which a gate voltage is attached to the normal layer. We demonstrate the penetration of the spin current and the spin-transfer torque into the right ferromagnetic region and show that, unlike graphene spin-valve junction, the spin-transfer torque in TI is very sensitive to the chemical potential of the NM region.
As an important result, by changing the chemical potential of the NM spacer and magnetization directions, one can control all components of the STT. It is interesting to note that both the resonance spin current and the resonance spin-transfer torque appear for the energies determined from the equation $k_x^NL=2n\pi$, where $k_x^N$ and $L$ are the wavevector and width of the NM region, respectively.
By increasing the chemical potential of the NM spacer, the amplitude of the STTs decreases while at large chemical potentials of $\mu_N$ there are intervals of chemical potential in which both the spin current and the spin-transfer torque become zero.
Moreover, we find that the spin-transfer torques versus the thickness of the central NM layer, display rapid oscillations as a function of the normal TI width. It is easily seen that for each configuration, the amplitude of the STT oscillations decays with the spacer width. The magnitude of the STT oscillations decreases as the chemical potential of the NM region increases.
These findings could open new perspectives for applications in spin-transfer torque magnetic random access memory (STT-MRAM) devices based on TI.

\end{document}